Miscalibration of simulations: A comment on Luebbert and Pachter (2024) "Miscalibration of the honeybee odometer" (arXiv.2405.12998v1)

Geoffrey W.Stuart, School of Psychological Sciences, University of Melbourne, Australia and School of Psychological Sciences, Macquarie University, Australia.

In this commentary I review the claim by Luebbert and Pachter (2024), henceforth "L&P", that the reported $R^2$ value in Srinivasan et al. (2000) was too high to be consistent with the reported means and standard deviations in the latter study. There is one serious limitation of the simulations conducted by L&P, and two flaws that compromise their findings. The reported $R^2$ value of Srinivasan et al. (2000) is within the expected range, as far as that can be determined given the limitations of the available data.

*Limitation*

It is not possible to accurately simulate the statistical analysis of Srinivasan et al. (2000) without access to the raw data (which is no longer available due to the passage of time). This is because the data structure corresponds to a multi-level model, a form of mixed effect model (e.g., Kreft & de Leeuw, 1998). There are four levels, in descending order (i) hive (ii) bee (iii) dance (iv) waggle. The only information reported in the paper concerns the hive that the bees came from, the total number of waggles analysed at each distance from the hive, and the pooled standard deviation for the full complement of waggles at each distance (Table 1 of Srinivasan et al., 2000). At most distances the mean waggle duration was calculated for a single hive. Importantly, the relevant mean waggle durations were pooled over the levels of bee and dance. While the total number of bees and dances was reported, the rest of the hierarchical structure was not specified. In particular, it is not known if waggle durations at different distances were obtained from the same bee. Therefore, any simulation must assume that the reported means are based on independent observations. A proper analysis, using mixed effect modelling, would incorporate various sources of covariance, such as for waggle durations measured using the same or different bees from a given hive, within the same or different dance, or at different food source distances for the same or different bees. These covariances would affect the expected range of $R^2$ values. It is therefore not possible to determine with full accuracy whether the reported $R^2$ is inconsistent with the reported pooled means and variances.

*Flaw 1*

In the first simulation, the level of analysis is at the second level in the hierarchical data structure, the mean waggle duration for each bee at each distance. These means are actually "means of means". To obtain them, the first step would be to calculate the mean waggle duration within each dance, noting the number of waggles involved, which could be different. Then, the mean waggle duration for each bee would be calculated for each dance performed by that bee, noting the number of dances. The grand means at each duration, weighted for the numbers at lower levels, would be the same as simply averaging waggle durations at the bottom level of the data hierarchy. However, the variances would not be the same, as they depend on the covariance structure within and between levels.

It is very important to note that the total number of waggles in Table 1 of Srinivasan et al. (2000) was used to calculate the mean waggle duration at each distance. The reported standard deviations are aggregated over all levels and are for all waggles – the lowest level in the data hierarchy. These standard deviations are not the standard *errors* that apply to the *mean* waggle durations for each bee. This leads to the main flaw in the first simulation. The simulated mean waggle duration for each bee (at the second highest level in the data hierarchy) was sampled from the variance estimate (standard deviation) at the lowest level (individual waggles). Not only that, the number of observations at the second highest level was simulated as the *number of bees*. Obviously, this is incorrect as the number of waggles should have been used. L&P, by using this method, have inflated the standard error of the mean waggle duration. By using only one simulated observation for each bee, *which is not consistent with the actual experiment*, they have confused the standard error of a mean with the standard deviation of the observations used to calculate that mean. This is a common confusion (Cumming, Fidler & Vaux, 2007) but it is critical to their claims.

The relationship between the standard error of the mean and the standard deviation is given by a well-known formula that can be found in any introductory statistics textbook:

Standard error = Standard deviation / $\sqrt{n}$ , where *n* is the number observations

Note that the standard error of the mean is equivalent to the standard deviation of the observations only in the case where *n*=1, which was not the case in Srinivasan et al. (2000) or any previous or subsequent study, all of which averaged over many individual waggles. By taking only one observation per bee from the standard deviation of the overall waggle distribution at each food source distance, L&P have implicitly assumed that the sole source of variance at the lowest level of the data hierarchy is that between bees, and that there is no variance between dances or waggles that could be averaged out to improve the estimate of mean waggle duration per bee. It is well known that there is variation between the duration of waggles within the same dance of the same bee (e.g., Schürch et al. 2013). That is why all researchers since von Frisch have averaged over a number of waggles to estimate the relationship between waggle duration and distance to a food source.

The effect of this error by L&P is to markedly inflate the variance of mean waggle duration between bees. If *n* represents the number of waggles contributing to an estimate of mean waggle duration, and *k* represents the number of bees who performed the total *n* waggles, the standard error of the mean waggle duration assuming only one waggle per bee is SD/$\sqrt{k}$ rather than SD/$\sqrt{n}$ . This inflation of the simulated variance severely attenuates the $R^2$ values obtained by when regressing mean waggle duration on distance. This is what is seen in the simulated results, in which nearly all of the simulated values are lower than the originally reported $R^2$ value (Figure 6, left panel of L&P). This is an artifact of the flawed simulation and therefore says nothing about the credibility of the $R^2$ value reported by Srinivasan et al. (2000).

*Flaw 2*

The second simulation, relegated to Supplementary material, does not suffer from Flaw 1. However, it is incorrectly asserted that this simulation assumes that there is no variation in the mean waggle duration between different bees — with the implication that this supposedly incorrect assumption is the explanation for the more conservative result. In fact, the variance for the pooled set of waggle durations represents all sources of variance, within bees, between bees, within dances and between dances. It also includes any effect of quantisation due to limited video frame rates, so there is no necessity to simulate that again. Notably, in this simulation, the proportion of $R^2$ values that exceed the original value is 19% (Supplementary

Figure 3 of L&P). This is not strong evidence against the credibility of the original reported $R^2$ of Srinivasan et al. (2000).

However, due to a second flaw in both simulations, the expected proportion of simulated $R^2$ values exceeding the reported value is still an underestimate. The simulations have confused the observed sample means with assumed population means. Every study of the bee "odometer" has concluded that the underlying relationship is linear (see Figure 2 of L&P). The statistically unbiased estimate of the underlying function prior to simulation is the observed regression line. As a sample estimate, the estimated line is subject to statistical error, quantified by the standard errors for the slope and intercept of the regression line. It is not expected that any true replication would yield the same slope, intercept or observed mean waggle duration at each distance. Nonetheless, the regression line is the proper baseline for judging the credibility of the observed scatter around it. By using the observed means as the starting point of their simulations, L&P have "baked in" the original sampling error of Srinivasan et al. (2000), and then added further sampling error based on the variance of waggle duration. This will attenuate a proportion of simulated $R^2$ values, and expand the variance. A valid test of the credibility of the originally reported $R^2$ value should have used the expected values from the regression equation, which lie on a straight line, as the starting point.

When the expected values are substituted for the original observed means in the second simulation, using the Python code of L&P, the percentage of simulated values exceeding the original $R^2$ rises to 96.67% (see Figure 1). In conclusion, there is no evidence that the high $R^2$ of Srinivasan et al. (2000), or indeed any of the other studies listed in Table 2 of L&P, do not reflect the analysis of genuine data.

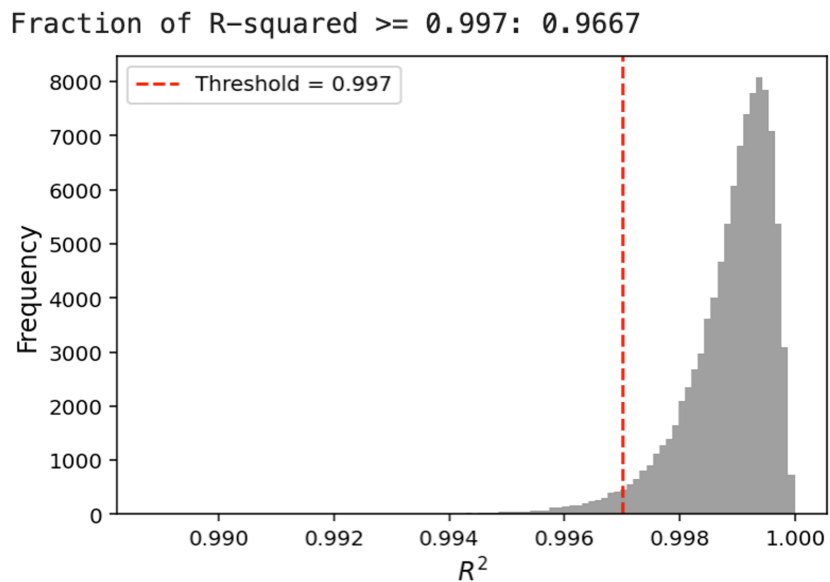

Figure 1.

Table 1 summarises the effect of these two errors, alone or in combination, on the distribution of $R^2$ values from different simulations using the original or slightly modified Python code of L&P. The percentage of simulated $R^2$ values that exceed that reported by Srinivasan et al. (2000) varies from 0.53% to 96.67%. The latter value is subject to the independence assumption, which is violated, but to an unknown degree. However, the assumption of complete independence is almost certainly closer to the true effective degrees of freedom. In the first simulation of L&P it was falsely assumed that there is zero variance due to dance or waggle, i.e. (i) perfect dependence at those levels, and (ii) an effective degrees of freedom that is the number of bees. As noted above, this does not follow the analytic strategy used by Srinivasan et al. (2000).

Table 1. *Percentage of simulated $R^2$ values exceeding 0.997, as reported by Srinivasan et al. (2000), (i) when bee or waggle is used as the unit of analysis and (ii) when observed means are used instead of predicted values from a linear regression line. All simulations were performed using Python code of Luebbert and Pachter, with predicted regression means substituted for observed means where applicable. Note that L&P did not seed their random number generator, so values change slightly over different runs. Ticks and crosses represent correct and incorrect assumptions of simulations.*

| Simulation Baseline | Replicate | |
|---|---|---|
| | Bees ✘ | Waggles ✓ |
| Observed Means ✘ | 0.53% | 19.53% |
| Regression Line ✓ | 0.92% | 96.67% |

*Reported $R^2$ values from other studies.*

Since it is not possible to simulate the study of Srinivasan et al. (2000) with the required level of accuracy, the results of similar experimental studies are critical. Schürch et al (2013) estimated the slope and intercept of the regression line relating waggle duration to distance to a food source. They used mixed effect modelling to analyse their data, overcoming the limitation of studies that only report and analyse mean waggle duration. Despite considerable variation in individual waggle durations, the reported $R^2$ for a linear regression relating mean waggle duration to distance to a food source was 0.9726. They reported that the linear $R^2$ value for the original study of von Frisch was 0.9906. These values are just under the value of 0.997 reported by Srinivasan et al. (2000). Without the ability to calculate a confidence interval for the $R^2$ value reported by Srinivasan et al. (2000) it is not possible to infer that these values differ significantly. There is thefore no proof of "potential manipulation" by Srinivasan et al. (2000). In all of these experimental studies the $R^2$ values reflect the precision obtained by the simple expedient of averaging the duration of a large number of waggles prior to regression. They are therefore not surprising or indicative of anything untoward in the original data.

*References*

Cumming G, Fidler F, Vaux DL. (2007). Error bars in experimental biology. Journal of Cell Biology, 177(1):7-11. doi:10.1083/jcb.200611141

Kreft I, de Leeuw J. (1998). Introducing Multilevel Modeling. Sage Publications, Inc. doi: 10.4135/9781849209366

Luebbert L, Pachter L. (2024). The miscalibration of the honeybee odometer. https://doi.org/10.48550/arXiv.2405.12998

Schürch R, Couvillon MJ, Burns DDR, Tasman K, Waxman D, Ratnicks FLW. (2013) Incorporating variability in honeybee waggle dance decoding improves the mapping of communicated resource locations. Journal of Comparative Physiology A, 199:1143-1152. doi: 10.1007/s00359-013-0860-4

Srinivasan MV, Zhang S, Altwein M, Tautz J. (2000) Honeybee navigation: Nature and calibration of the "odometer". Science, 287(5454): 851–853.
doi:10.1126/science.287.5454.851